# Phase transition from focal conic to cubic smectic blue phase in partially fluorinated cyano-phenyl alkyl benzoate ester doped with ultrahigh twisting power chiral dopant

Prasenjit Nayek,[1] Sebastian Gauza,[3] Guoqiang Li[1,2,*]

[1]Department of Ophthalmology and Visual Science, The Ohio State University,

1330 Kinnear Road., Columbus, OH 43212, USA

[2]Department of Electrical and Computer Engineering, The Ohio State University,

1330 Kinnear Road., Columbus, OH 43212, USA

[3]LC MATTER Corporation, 4000 Central Florida Blvd., Orlando, FL 32816

[4]Department of Physics, University of Colorado, 2000 Colorado Ave., Boulder, CO 80309

*Corresponding author: li.3090@osu.edu

Blue phase liquid crystal (BPLC) has important applications in adaptive lenses and phase modulators due to its polarization-independent property. During our efforts for development of the new materials, we found a novel phenomenology of phase transition, from focal conic smectic to smectic blue phase in a partially fluorinated cyanophenyl alkyl benzoate ester based nematic liquid crystal (LCM-5773) doped by ultra-high twisting power [H.T.P~160 $\mu m^{-1}$] chiral dopant (R5011/3 wt%). Polarized optical microscopy (POM) investigations revealed focal conic and fan-shaped textures typical for columnar mesophases. These focal conic domains (FCDs) are squeezed under electric field and finally at a critical electric field they undergo a dark state. When the electric field is withdrawn, the FCDs are regrown in a one dimensional array with smaller domain size. Interestingly, we have observed the domain size of the FCDs can grow several times by decreasing the cooling rate (0.02 $^{o}$C/min.) ten times without any change in the phase sequence. In blue phase (BP), we have observed curved platelet texture and grain boundaries filled by small platelets, which is completely different from conventional cholesteric BP. The blue phase platelet size (PLS) also increases significantly at low cooling rates. The thermal control of FCD and PLS size has increasing demand for the construction of devices with optimal performances.

**Introduction**

Liquid crystal (LC) is ubiquitous and embedded to our daily life through user-friendly devices.[1] These devices can be categorized as display and non-display devices.[2] The latter includes phase modulators, adaptive lenses, temperature sensors, biosensors, etc. The LC adaptive lenses[3-6] have the advantage in tuning the focusing power by applying a low voltage. They may have very high impact in the field of vision care, especially for correction of prebyopic eyes that need add power for near-vision tasks due to aging of the eyes. It is a natural phenomenon and eventually affects almost everybody, so this application has a broad market. We have developed several prototypes for this application.[3-6] The currently available devices use nematic LC materials, which is polarization dependent, and therefore two devices with orthogonal buffering directions are needed to be integrated together to form a polarization-independent LC adaptive lens. Blue phase (BP) LC material has polarization-independent electro-optic response and has the potential to greatly simplify the structure of the LC lenses and phase modulators. Therefore it is important to develop new BPLC materials that are suitable for this application. The new finding to be reported in this paper is the intermediate result of our efforts towards this goal.

In LC systems most non-trivial behaviors originate in the form of frustration produced by two or more conflicting controlling parameters.[7-9] Frustrations can generate novel phase transition between ordered phases and complex spatial structures at the cost of minimization of local conflict. Among the frustrated LC systems, twist grain boundary (TGB) phase is unique and it arises from the frustration of chirality combined with a smectic density wave. TGB phase was predicted theoretically by de Gennes,[10] and Renn and Lubensky[11] as the LC analog of the Abrikosov type II superconductor and has been found experimentally in chiral thermotropic LC by Goodby *et al*[12,13] The TGB phase generally arises at the phase transition from the isotropic or chiral nematic (N*) to SmA or SmC* phases[14,15] and the temperature range is very short.[16,17]

However, recent investigations found wide temperature ranged TGB phases.[18] Further study on those materials revealed that a large tilt susceptibility and a large penetration length for tilt of the director, are responsible for bulk giant smectic blocks.[18] Another example of chiral frustrated phase is cholesteric blue phase (BP) LC. In cholesteric BP, the lowest energy director configuration consists of double-twist cylinders in which the director rotates spatially about any radius of a cylinder.[19] These double-twist cylinders are incompatible with the requirement of continuity, sodisclinations arise where the cylinder directors are mismatched which are essential for relieving the elastic strain energy. Among the three types of stable BPs, two of them, BP I and BP II, exhibit cubic symmetry in which the orientational order is periodic and long range in three dimensions. BP I has a body-centered cubic structure, whereas BP II has a simple cubic structure. The BP I director field possesses the symmetry group $O^8$ ($I4_132$) and BP II, $O^2$ ($P4_232$).[20] The symmetry of BP III phase is the same as that of the isotropic phase.[21] On the other hand, cholesteric BP arises between isotropic and N* phase with very short range except in a few cases where a SmA to BP I transition has been observed.[22-24] Li $et\ al$[25,26] reported the observation of a different phase sequence I-BPs-TGBA-TGBC-SmC* (I indicates the isotropic phase) in a fluoro-substituted chiral tolane derivative. Many groups tried to solve the main impediment of short temperature of this delicate phase by different techniques.[27-29] Recently new BPs have been discovered in a chiral material (called FH/FH/HH-n BTMHC, where n indicates the paraffinic chain length) with the following phase sequence: TGB phases–BP–Iso.[30-32] In this sequence, three BPs arise which are not conventional BP rather termed as smectic BP ($BP_{Sm}$). Quasi long range smectic order has been established by x-ray scattering experiment. Thus smectic BPs are doubly frustrated systems where the chirality extends in the three spatial dimensions like classical BPs, and the helical twist competes with smectic order like for TGB phases. There have been reported three $BP_{Sm}$ phases namely, $BP_{Sm}1$, $BP_{Sm}2$ and $BP_{Sm}3$. The orientational symmetries of the smectic BPs are cubic for $BP_{Sm}1$, hexagonal for $BP_{Sm}2$, and isotropic for $BP_{Sm}3$, respectively.[33] Yamamoto *et al* [34] reported another new, optically isotropic smectic BP ($SmBP_{iso}$) which can appear in the phase sequence of smectic BPs by mixing chiral monomer (3B1M7) and its twin (BMHBOP-6). Interestingly, they have observed six modulated phases including three different SmBPs (SmBPX1, SmBPX2 and SmBPX3) whereas host monomers had only isotropic liquid (Iso) and smectic-A ($S_A$) phases. Apart from the conventional frustration on chiral LC system, recently Araki *et al* [35] reported progress of topological frustration in nematic LC confined in porous material which has memory effect owing to topology of the confining surface that maximizes in a simple periodic bicontinuous cubic structure. Xu *et al* [36] reported soluble, chemically oxidized graphene or graphene oxide sheets which can form chiral LCs in a TGB phase-like model with simultaneous lamellar ordering and long-range helical frustrations.

In our present study, we are reporting a new phenomenology of phase transition, from focal conic (smectic) to cubic smectic BP in partially fluorinated cyanophenyl alkyl benzoate ester doped by ultra-high twisting power chiral dopant during cooling the cell. Up to now to our knowledge, there is no intermediate phase reported in conventional BPs and/or Smectic BPs. Instead of platelet $BP_{Sm}2$ or amorphous $BP_{Sm}3$, we observed focal conic texture. These focal conic domains (FCDs) ruptured to optically isotropic state under a critical electric field and they reappear with one dimensional array with smaller domain size when the electric field is withdrawn without any alignment layer on electrodes. Interestingly, we also observed the domain size of the FCDs can grow several times by changing the cooling rate and the transition sequence remains unchanged. Also unique grain boundary has been observed where grains are surrounded by colored platelets oriented in a ropelike manner. The blue phase texture is quite different from conventional cholesteric BP. Curvature in platelet and twist is observed, which is quite surprising.

**Experimental**

The host nematic LC used, LCM-5773, has the temperature interval from - 40$^o$C up to the transition to the isotropic phase at T$_{NI}$ = 70 $^o$C. LCM-5773 has a birefringence, Δn = 0.23 (at 20$^o$C and 589 nm), and dielectric anisotropy, Δε = 67 (positive anisotropy). The optically isotropic phase was induced by mixing chiral dopant R5011 (3 wt%) with the host NLC, LCM-5773. The LC cells were fabricated by assembling two 1-inch square glass substrates uniformly coated with indium tin oxide (ITO) electrode film. The 5μm silica spacers dispersed in ethanol were drop casted onto one glass substrate and spin coated at 3000 rpm for 30 sec using Laurell WS-650-23B spin coater. During the spin coating, the evaporation of the ethanol takes place. The spacers coated glass substrate is then carefully placed onto the bare glass substrate and glued together using epoxy resin sealant dispersed with 5 μm spacers. The mixture was injected into cell and then thermally heated up to 70 $^o$C by the Instec-mK1000 temperature controller with stage, HCS302. The filled cells were cooled down to room temperature at different cooling rates (0.02- 0.2 $^o$C/min.). The textures were taken by a polarized optical microscope (POM) Leitz Laborlux 12 pol. fitted with a Nikon digital sight DS-Fi2 digital camera. The input voltage of the cell was applied by a function generator (Agilent 33220A).

**Results and discussion**

Figure 1(a) shows the general chemical structure of the host nematic LC (LCM-5773). Fluoro group provides an excellent resistivity modest dipole moment and low viscosity.[37-39] Figure 1 (b) shows the chemical structure of the chiral dopant R5011. The helical twisting power of the chiral dopant R5011 is >160. We have used the LCM-5773 mixed with 3 wt% of chiral dopant R5011. The textures were studied by injecting the mixture into a 5 μm-thick ITO coated LC cell during cooling. The POM textures are depicted in Fig. 2 at a cooling rate 0.02 $^{o}$C/min. At 66.2$^{o}$C we can see focal conic texture. Focal conics are oriented arbitrarily. At 66.5 $^{o}$C (Fig. 2(b)) focal conics grows larger in size (average ~ 80 μm) and some superimposes with each other. Focal conic textures are the indicator for smectic phase. Figure 2 (c) shows the transition from the focal conic to the BP at 65.2$^{o}$C. Further cooling enables more prominent platelet structures which are shown in Figs. 2 (d) - (f). Figure 2 (g) shows the transition from the BP to the TGB phase. The grain boundaries are clear at lower temperatures as shown in Figs. 2 (h) - (i). At high sample cooling rates (rate 0.2$^{o}$C/min), we observed the texture size gets smaller for focal conic as well as BP. We have observed little shift of the transition temperatures, as shown in the Fig. 3. Figure 3 (a) shows the focal conic texture with smaller size at 68$^{o}$C with respect to the previous cooling rate. Figure 3 (b) shows the starting transition from smectic to BP at 67$^{o}$C. BP platelet textures are more prominent in 66.6 $^{o}$C (Fig. 3(c)) and at 65.5 $^{o}$C (Fig. 3(d)). Again, the transition from BP to TGB phase took place at 65$^{o}$C (Fig. 3(e)). It is worthy to mention that transition from smectic focal conic to BP LC is not normal. In general, for previous report of smectic blue phases, there is no evidence where focal conic texture arose before BP during cooling from isotropic liquid phase. Therefore this transition is novel. According to the phase sequence of smectic BP, hexagonal BP$_{Sm}$2 may arise before cubic BP$_{Sm}$1, but the texture is mosaic, contrary to our observation.[31] In this case focal conic/fan shaped texture indicates the phase may be hexagonal, but why we cannot see mosaic texture is yet to be discovered. It seems that smectic BPs are not only different from structural point of view, but also their phase sequences are innovative. Here we have used a highly dielectric anisotropic host with laterally substituted fluorine atoms which may have unique structure, strong dipole moment and manifest such type of transition. Also the chiral dopant used is not simple in structure. It has a little tuning fork (short arms). Twisting power is also exceptionally high which may be one of the reason for such structures because previously many authors used different shape of molecules for the synthesis of different kinds of BPs.[40,41] This phase sequence is completely different from previously reported cholesteric BP phase sequence where BP arises in between isotropic state and chiral nematic phase. Also the blue phase which transforms from smectic should be smectic BP as there is no chiral nematic phase in the phase sequence. Careful observation reveals some unusual grain boundary disclination lines as depicted in Fig. 4 (by white dotted lines) depart from the conventional BPLC cholesteric textures. The magnified image in Fig. 4, left corner, is depicted and it is clear that at the grain boundary (the region between red arrows) different colored smaller platelets are arranged in curves that means inside the grain boundary BP unit cells have different orientation, schematically illustrated by different colored cubes. Inspired by the above texture we have cooled the sample for a second cycle and the textures seem to be more uniform and prominent as shown in Fig. 5. The most interesting textures observed are curved platelets which seem to have some twisting, as shown in Fig. 5 (a) (inset of Fig. 5 (b)). This type of structures has a chiral shape and consumes less space by packing in a twisted manner. As indicated by small arrows, possibly there are some twisting between successive curved platelets. Apart from that each isolated giant grains are surrounded by small platelets as shown in the Figs. 5 (c)-(d), by curved arcs. Grain boundaries are filled with smaller platelets and look like a curved arc. Figures 5(d) and (e) show a more interesting phenomenon that two grains are separated by two different rope-like grain boundaries. This type of BP texture is absolutely new and expected to be manifestation of smectic blue phases. Figures 6 (a)-(c) show that as the amplitude of the square wave voltage of frequency 1 kHz increases, the focal conics are squeezing and finally at a critical electric field the texture gets dark (Fig. 6 (d)). As we predicted that the focal conic texture may be hexagonal, here we are getting another clue that it might be hexagonal columnar type phase consisting of smectic cylinders, because under the vertical electric field the texture is going to dark state when the sample is filled in a sandwiched glass cell. This kind of texture and vertical electric field induced transition have been reported by other authors [42, 43] and are similar to our observations. Figure 7(a) shows the texture at 25$^o$C during cooling and reveals grain boundary which is identified as TGB phase. Texture color, and electric field induced switching indicate the phase may be TGBC*. Under an electric field the texture changes its color slightly due to electro clinic effect and switching is observed under POM as shown in Fig. 7 (b).

**Conclusion**

The present findings pose new theoretical challenges and potentially open the way for smectic blue phase liquid crystal research. Due to the complicated coupling of the coexisting smectic layer order and helix, the internal structure and the stabilization mechanism of the SmBPs have not yet been fully investigated. In our sample cooling from isotropic phase, the sample may come to a columnar phase, then it transform to smectic cubic blue phase, and finally to TGB phase. This columnar phase seems to be double twisted. In TGB phase, first TGBA* phase arose, then TGBC, and finally it came to TGBC* phase. The temperature ranges are as follows from texture observation Iso-67.13$^o$C-FCD-65.75$^o$C-BP-63.4$^o$C-TGBA-60$^o$C-TGBC*-25$^o$C, during cooling at 0.02$^o$C/min. Although further details of the structural observation are yet to be produced for more clear understanding of the phase, the molecular shapes, length, twisting power are imperative factors for the origin of new phases and structures in the field of soft matter.

**Author Contributions**

The manuscript was written through contributions of all authors. All authors have given approval to the final version of the manuscript.

**Acknowledgments**

G. Li would like to thank the support from National Institutes of Health National Eye Institute (through grant R01 EY020641), National Institute of Biomedical Imaging and Bioengineering (through grant R21 EB008857), National Institute of General Medical Sciences (through grant R21 RR026254/R21 GM103439), and Wallace H. Coulter Foundation Career Award (through grant WCF0086TN). G. Li also thanks Varun Penmatsa for help with assembly of the LC cells.

**Figure caption:**

Fig. 1. Schematic chemical structure of (a) host NLC used (general structure), (b) chiral dopant R5011.

Fig. 2. The texture during cooling at a cooling rate 0.02 °C/min. (a), (b), the focal conic texture; (c), the transition to BP phase; (d)-(f), the BP phase; (g), transition to TGB phase; (h)-(i), TGB phase.

Fig. 3. The POM texture during cooling at a cooling rate 0.2 °C/min. (a), the focal conic texture; (b)-(c), the transition to BP phase; (d)-(e), the BP phase; (f), transition to TGB phase; (g)-(i), TGB phase.

Fig. 4. BP texture under POM at 63.8 °C. White dotted lines represent the grain and the lines represent declination around the grains. Inset picture clearly shows one grain and grain boundary (by red arrows) where platelets are arranged one dimensionally along the grain boundary. Different color represents different orientations of the cubes as depicted schematically by colored cubes.

Fig. 5. The BP texture under POM, at 63.5 °C during the second cycle of the cooling ((b)). Inset (a) represents the unusual type of platelet shape (curved) and probably there is some twist among the platelets; (c), a giant grain and the grain boundary (GB). Red arrows and yellow arc show the GB orientation; (d), the different grains and GBs. Interestingly, individual grains are surrounded by GBs and GB consists of one dimensional curved platelets; (e), the two grains separated by two different GBs.

Fig. 6. Focal conic texture under POM at 66 °C. (a), without any voltage; (b), with 7 $V_{pp}$ and 1kHz square voltage applied; (c), with 10 $V_{pp}$ and the same frequency; (d), 15 $V_{pp}$ shows dark

texture;, (e), when the electric field is switched "OFF" from 15 $V_{pp}$; (f), after waiting 3 min. from the switching "OFF" condition. Lines represent the one dimensional array of focal conics.

Fig. 7. POM texture at 25 °C. (a) without any voltage applied. The grain and grain boundary of TGB phase are prominent; (b) with voltage 20 $V_{pp}$ and frequency 1 kHz square wave.

(a)
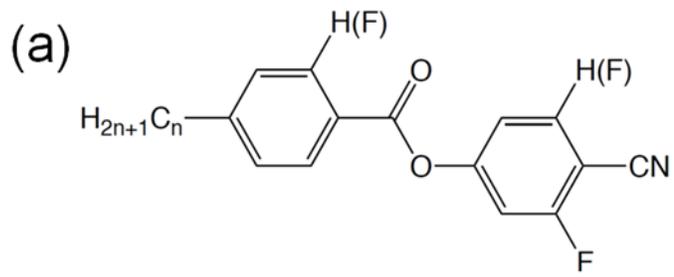

(b)
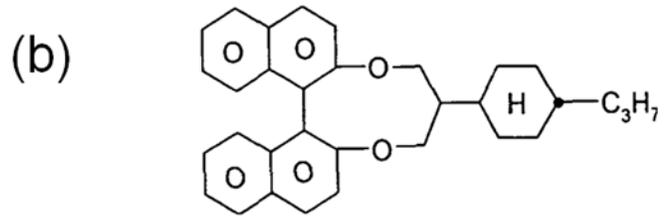

Fig.1

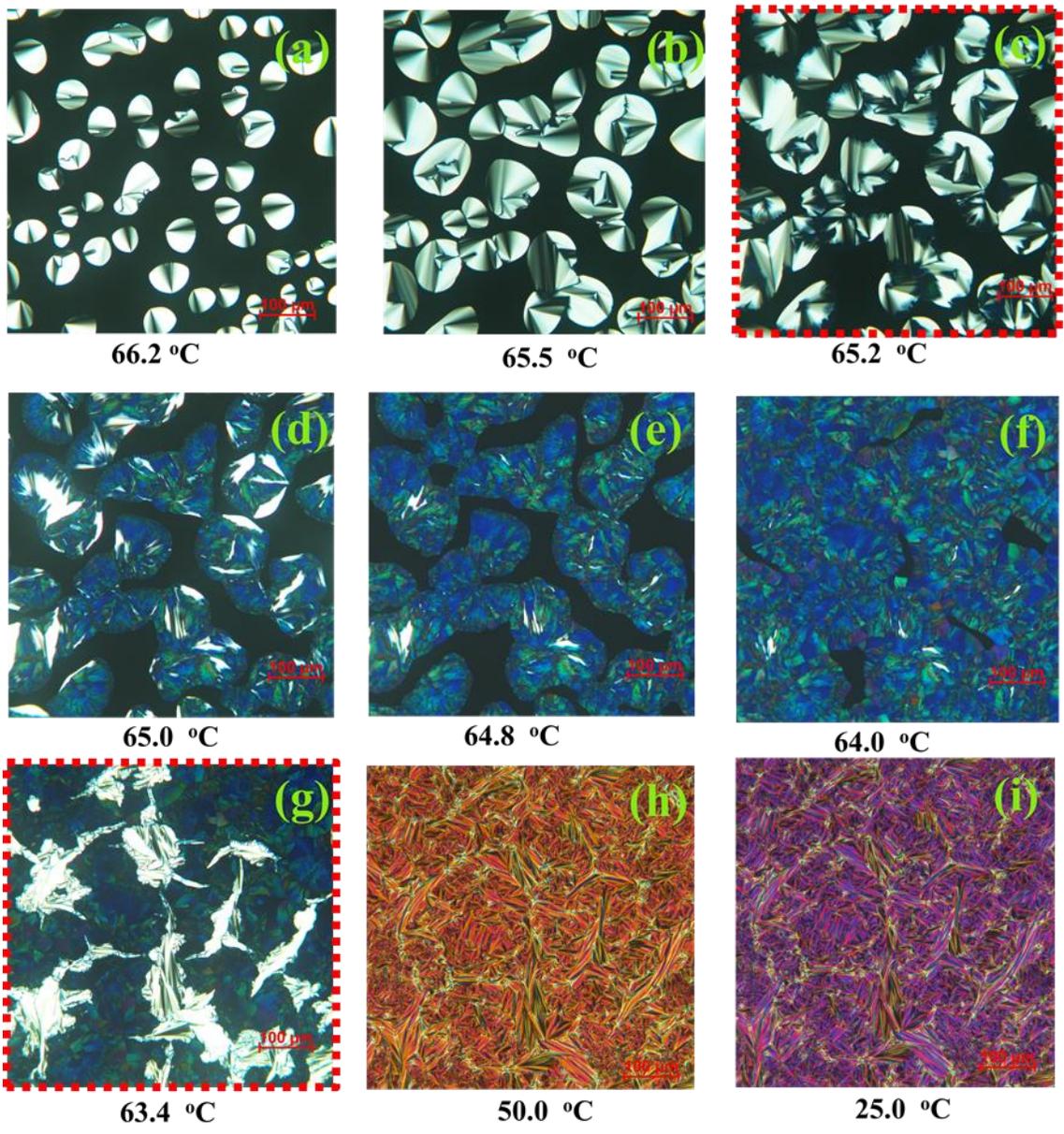

Fig.2

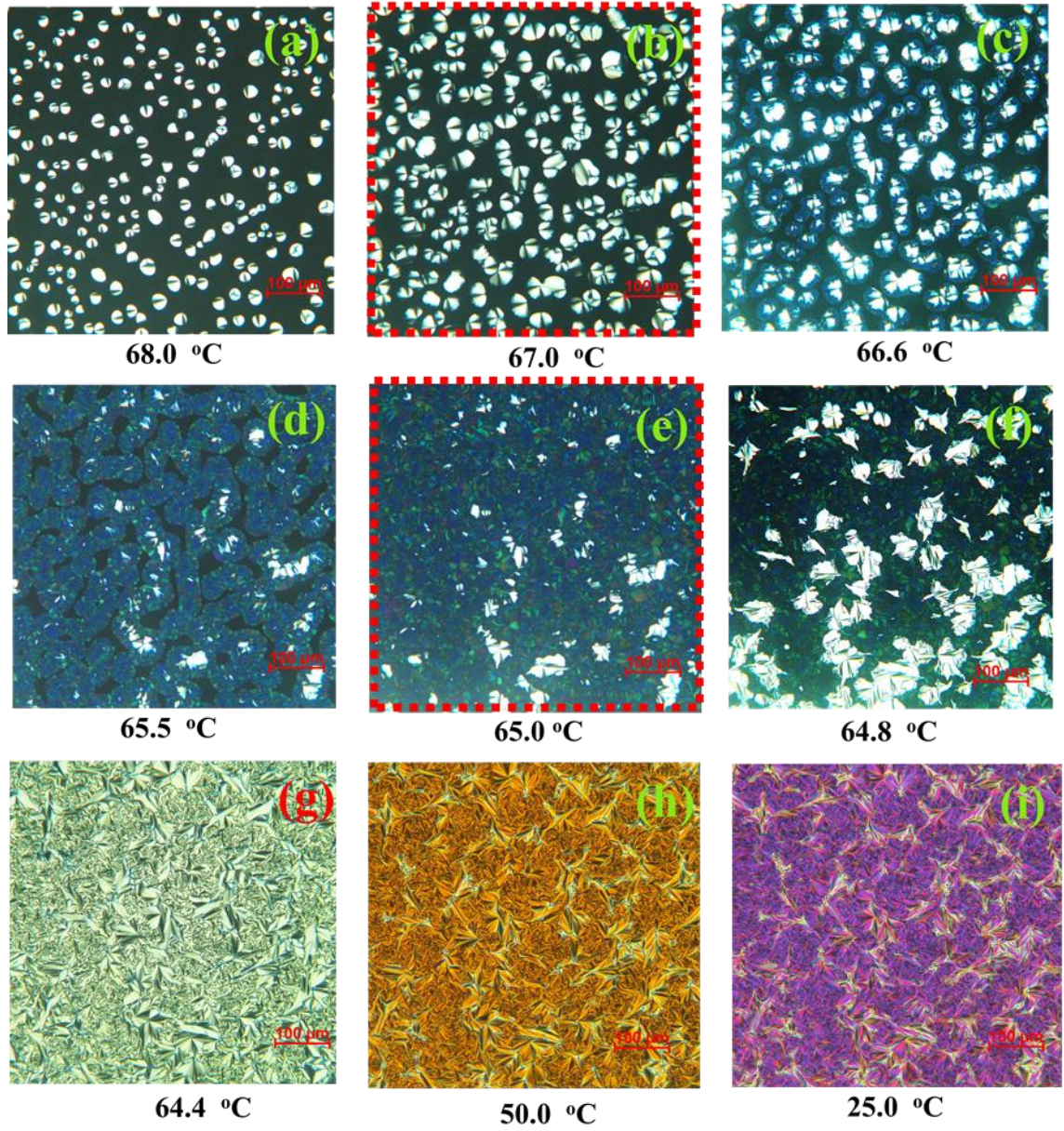

Fig.3

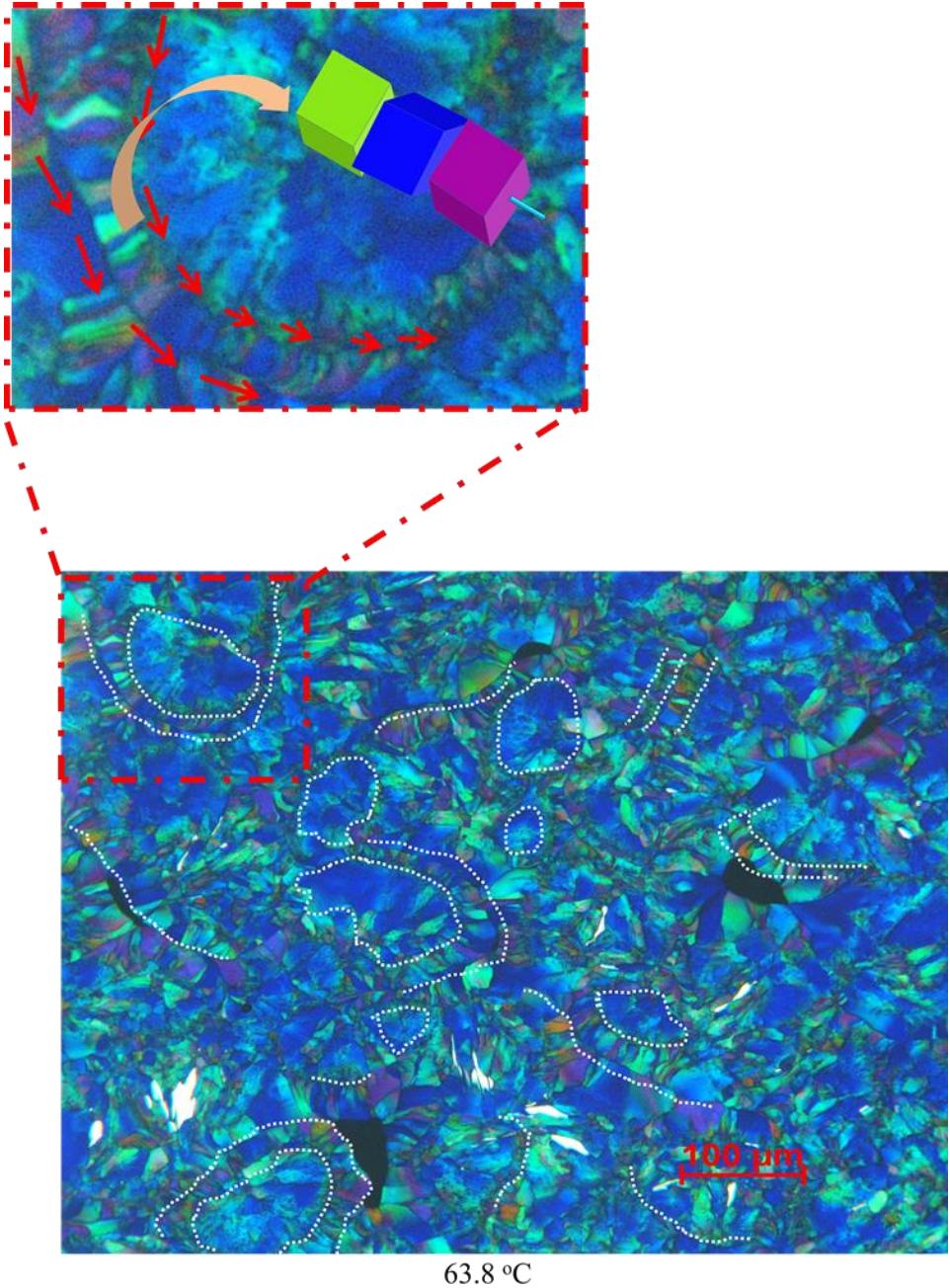

63.8 ºC

Fig.4

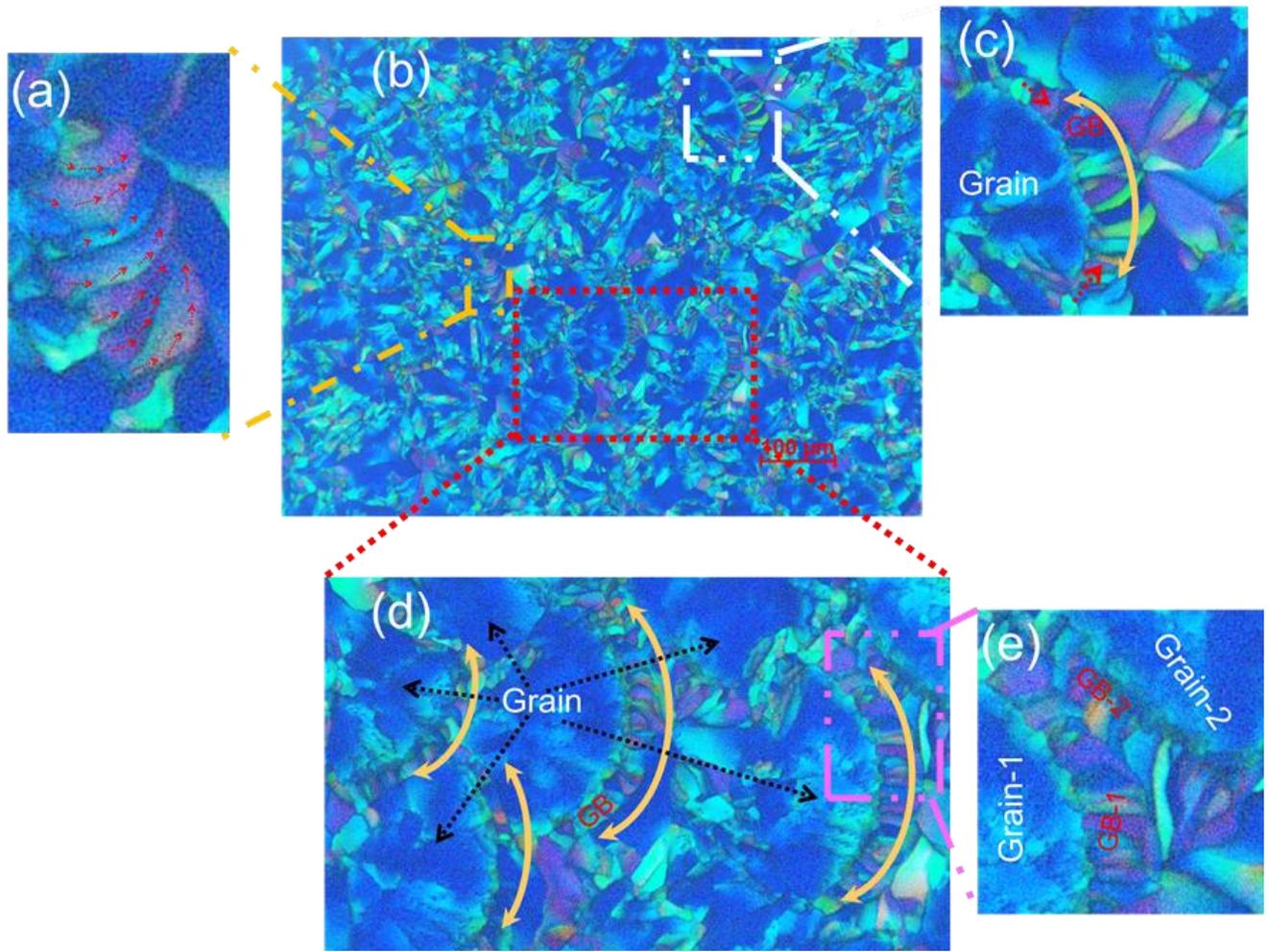

Fig.5

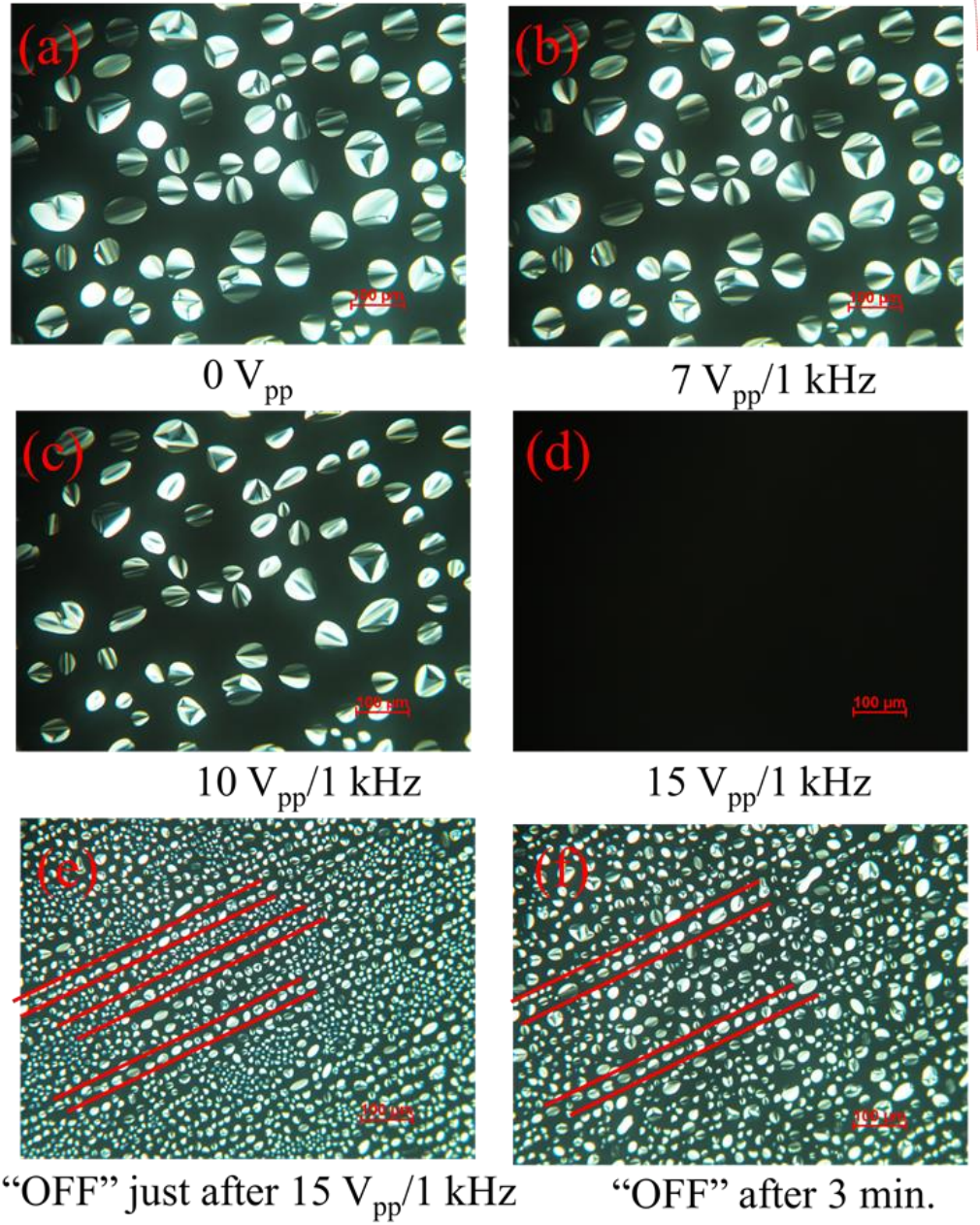

Fig.6

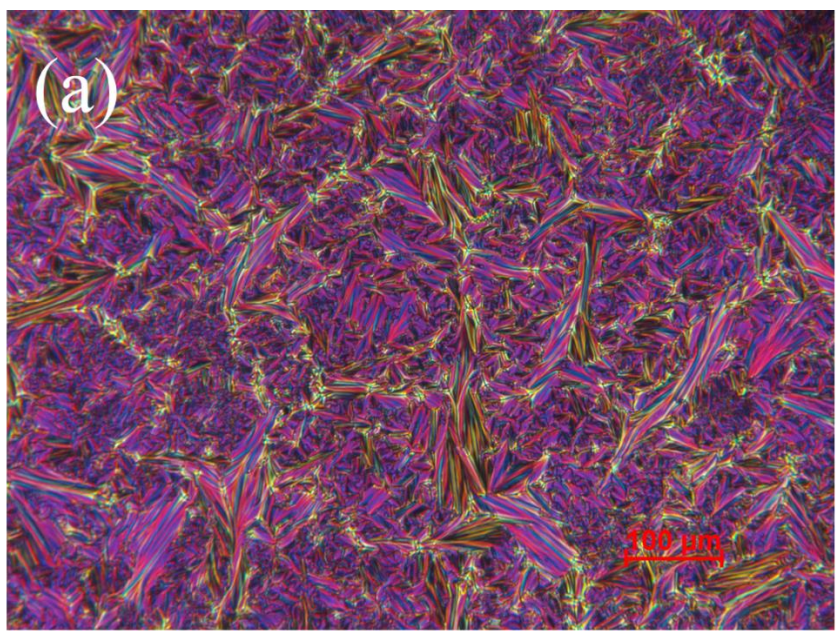

25 °C/ 0 V$_{pp}$

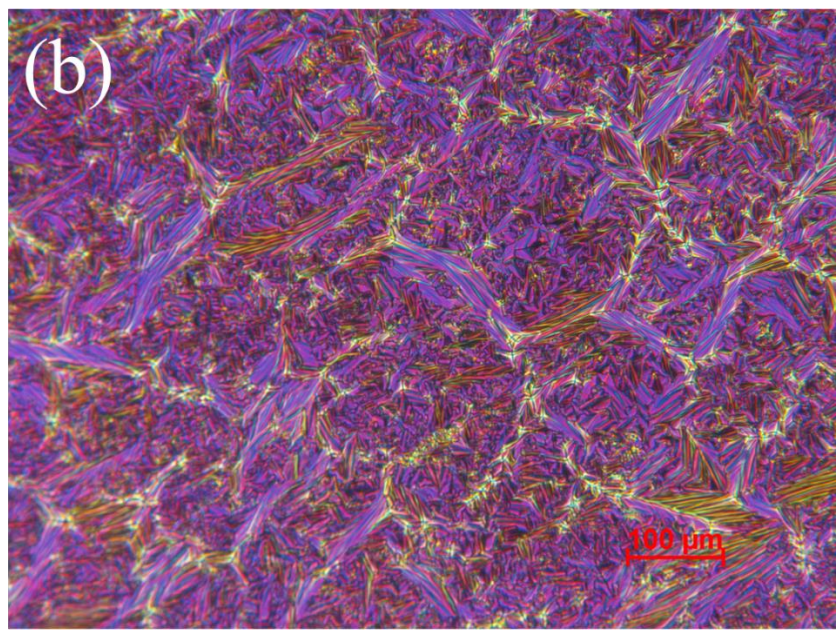

25 °C/ 20 V$_{pp}$

Fig.7